\documentclass[12pt]{article}
\usepackage[all]{xy}
\usepackage[dvips]{graphicx}
\usepackage{enumerate}
\usepackage{graphics}

\usepackage{amsmath}
\usepackage{bm}

\begin{document}
\begin{center}
{\bf A VERSATILE INTEGRAL IN PHYSICS AND ASTRONOMY}\\
 \vskip.3cm A.M. Mathai\\
Centre for Mathematical Sciences,  Arunapuarm P.O., Pala,
Kerala-686574, India, and\\
Department of Mathematics and Statistics, McGill University, Canada\\
and\\
\vskip.3cm H.J. Haubold\\
Office for Outer Space Affairs, United Nations, Vienna International Centre, P.O. Box 500, A-1400 Vienna, Austria, and\\
Centre for Mathematical Sciences,  Arunapuarm P.O., Pala, Kerala-686574, India\\
\end{center}
\vskip.3cm \noindent {\bf Abstract} \vskip.3cm This paper deals with a general class of integrals, the particular cases of which are connected to outstanding problems in astronomy and physics. Reaction rate probability integrals in the theory of nuclear reaction rates, Kr\"atzel integrals in applied analysis, inverse Gaussian distribution, generalized type-1, type-2 and gamma families of distributions in statistical distribution theory, Tsallis statistics and Beck-Cohen superstatistics in statistical mechanics and the general pathway model are all shown to be connected to the integral under consideration. Representations of the integral in terms of generalized special functions such as Meijer's G-function and Fox's H-function are also pointed out.\\
\hrule \vskip.2cm \noindent Keywords: H-function, Kr\"atzel integral, generalized beta, gamma and inverse Gaussian densities, pathway model.\\
\vskip.2cm \noindent \hrule
 \vskip.3cm \noindent
{\bf 1.\hskip.3cm Introduction} \vskip.3cm In this paper we will consider a general class of integrals connected with the pathway model of Mathai (2005). These will enable us to address a wide class of problems in different areas such as inverse Gaussian processes in the area of stochastic processes, Kr\"atzel integral in applied analysis, generalized type-1, type-2 and gamma densities in statistical distribution theory, Tsallis statistics in non-extensive statistical mechanics, superstatistics in astrophysics problems, reaction probability integrals in nuclear reaction rate theory, and many related problems, which may be seen from the formalism introduced in this paper. Consider the following integral:
\begin{align}
f(z_2|z_1)&=\int_0^{\infty}x^{\gamma-1}[1+z_1^{\delta}(\alpha-1)x^{\delta}]^{-\frac{1}{\alpha-1}}
[1+z_2^{\rho}(\beta-1)x^{-\rho}]^{-\frac{1}{\beta-1}}\\
&\mbox{for  }\alpha >1,\beta >1,z_1\ge 0, z_2\ge 0,
\delta >0,\rho >0,\Re(\gamma+1)>0,\nonumber\\
&\Re(\frac{1}{\alpha-1}-\frac{\gamma+1}{\delta})>0,
\Re(\frac{1}{\beta-1}-\frac{1}{\rho})>0\nonumber\\
&=\int_0^{\infty}\frac{1}{x}f_1(x)f_2(\frac{z_2}{x}){\rm d}x
\end{align}
where $\Re(\cdot)$ denotes the real part of $(\cdot)$.
\begin{equation}
f_1(x)=x^{\gamma}[1+z_1^{\delta}(\alpha-1)x^{\delta}]^{-\frac{1}{\alpha-1}},
~f_2(x)=[1+(\beta-1)x^{\rho}]^{-\frac{1}{\beta-1}}
\end{equation}
with Mellin transforms
\begin{align}
M_{f_1}(s)&=[\delta z_1^{\gamma+s}
(\alpha-1)^{\frac{\gamma+s}{\delta}}]^{-1}
\frac{\Gamma(\frac{\gamma+s}{\delta})
\Gamma(\frac{1}{\alpha-1}-\frac{\gamma+s}{\delta})}{\Gamma(\frac{1}{\alpha-1})},\\
& \Re(\gamma+s)>0,
\Re\left(\frac{1}{\alpha-1}-\frac{\gamma+s}{\delta}\right)>0\nonumber\\
\intertext{and}\notag\\
M_{f_2}(s)&=[\rho(\beta-1)^{\frac{s}{\rho}}]^{-1}\frac{\Gamma(\frac{s}{\rho})
\Gamma(\frac{1}{\beta-1}-\frac{s}{\rho})}{\Gamma(\frac{1}{\beta-1})}\\
&\Re(s)>0,
\Re\left(\frac{1}{\alpha-1}-\frac{s}{\rho}\right)>0.\nonumber
\end{align}
Hence the Mellin transform of $f(z_2|z_1)$, as a function of $z_2$,
with parameter $s$ is the following:
\begin{align}
M_{f(z_2|z_1)}(s)&=M_{f_1}(s)M_{f_2}(s)\nonumber\\
&=
\frac{1}{\delta~z_1^{\gamma+s}(\alpha-1)^{\frac{\gamma+s}{\delta}}}
\frac{\Gamma(\frac{\gamma+s}{\delta})\Gamma(\frac{1}{\alpha-1}-\frac{\gamma+s}{\delta})}{\Gamma(\frac{1}{\alpha-1})}\nonumber\\
&\times
\frac{1}{\rho(\beta-1)^{\frac{s}{\rho}}}\frac{\Gamma(\frac{s}{\rho})
\Gamma(\frac{1}{\beta-1}-\frac{s}{\rho})}{\Gamma(\frac{1}{\beta-1})}\\
&\mbox{for  }\Re(\gamma+s)>0,
\Re(\frac{1}{\alpha-1}-\frac{\gamma+s}{\delta})>0,
\Re(s)>0,\nonumber\\
&\Re(\frac{1}{\beta-1}-\frac{s}{\rho})>0,z_1>0,z_2>0.\nonumber
\end{align}
Putting $y=\frac{1}{x}$ in (1) we have
\begin{equation}
f(z_1|z_2)=\int_0^{\infty}\frac{y^{-\gamma}}{y}[1+z_1^{\delta}(\alpha-1)y^{-\delta}]^{-\frac{1}{\alpha-1}}
[1+z_2^{\rho}(\beta-1)y^{\rho}]^{-\frac{1}{\beta-1}}{\rm d}y.
\end{equation}
Evaluating the Mellin transform of (7) with parameter $s$ and
treating it as a function of $z_1$, we have exactly the same
expression in (6). Hence
\begin{equation}
M_{f(z_2|z_1)}(s)=M_{f(z_1|z_2)}(s)= \mbox{right side in (6)}.
\end{equation}
By taking the inverse Mellin transform of $M_{f(z_2|z_1)}(s)$ one
can get the integral $f(z_2|z_1)$ as an H-function as follows:
\begin{equation}
f(z_2|z_1)=c^{-1}H_{2,2}^{2,2}\left[z_1z_2(\alpha-1)^{\frac{1}{\delta}}
(\beta-1)^{\frac{1}{\rho}}
\big\vert_{(\frac{\gamma}{\delta},\frac{1}{\delta}),(0,\frac{1}{\rho})}^{(1-\frac{1}{\alpha-1}
+\frac{\gamma}{\delta},\frac{1}{\delta}),(1-\frac{1}{\beta-1},\frac{1}{\rho})}\right]
\end{equation}
where
\begin{equation*}
c=\delta\rho z_1^{\gamma}(\alpha-1)^{\frac{\gamma}{\delta}},
\end{equation*}
where $H_{p,q}^{m,n}(\cdot)$ is a H-function which is defined as the
following Mellin-Barnes integral:
\begin{equation}
H^{m,n}_{p,q}\left[z\big\vert_{(b_1,\beta_1),...,(b_q,\beta_q)}^{(a_1,\alpha_1),...,(a_p,\alpha_p)}\right]=
\frac{1}{2\pi i}\int_L\phi(s){\rm d}s
\end{equation}
where
\begin{equation}
\phi(s)=\frac{\{\prod_{j=1}^m\Gamma(b_j+\beta_js)\}
\{\prod_{j=1}^n\Gamma(1-a_j-\alpha_js)\}}
{\{\prod_{j=m+1}^q\Gamma(1-b_j-\beta_js)\}\{\prod_{j=n+1}^p\Gamma(a_j+\alpha_js)\}}
\end{equation}
where $L$ is a suitable contour, $\alpha_j, j=1,...p, \beta_j,
j=1,...,q$ are real positive numbers, $b_j,j=1,...,q, a_j,
j=1,...,p$ are complex numbers and $L$ separates the poles of
$\Gamma(b_j+\beta_js), j=1,...m$ from those of
$\Gamma(1-a_j-\alpha_js), j=1,...,n$. For more details about the
theory and applications of H-function see Mathai and Saxena (1978) and Mathai et al. (2009).
\vskip.2cm The integral in (1) is connected to reaction rate
probability integral in nuclear reaction rate theory in the
non-resonant case, Tsallis statistics in non-extensive statistical
mechanics, superstatistics in astrophysics, generalized type-2,
type-1 beta and gamma families of densities and the density of a
product of two real positive random variables in statistical
literature, Kr\"atzel integrals in applied analysis, inverse
Gaussian distribution in stochastic processes and the like. Special
cases include a wide range of functions appearing in different
disciplines. \vskip.2cm Observe that $f_1(x)$ and $f_2(x)$ in (3),
multiplied by the appropriate normalizing constants can produce
statistical densities. Further, $f_1(x)$ and $f_2(x)$ are defined
for $-\infty <\alpha<\infty, -\infty<\beta<\infty$. When $\alpha >1$
and $z_1>0, \delta >0, f_1(x)$ multiplied by the normalizing
constant stays in the generalized type-2 beta family. When $\alpha
<1$, writing $\alpha-1=-(1-\alpha), \alpha <1$ the function $f_1(x)$
switches into a generalized type-1 beta family and when
$\alpha\rightarrow 1$,
\begin{equation}
\lim_{\alpha\rightarrow 1}f_1(x)={\rm e}^{-z_1^{\delta}x^{\delta}}
\end{equation}
and hence $f_1(x)$ goes into a generalized gamma family. Similar is
the behavior of $f_2(x)$ when $\beta$ ranges from $-\infty$ to
$\infty$. Thus the parameters $\alpha$ and $\beta$ create pathways
to switch into different functional forms or different families of
functions. Hence we will call $\alpha$ and $\beta$ pathway
parameters in this case. Let us look into some interesting special
cases. Take the special case $\beta\rightarrow 1$,
\begin{align}
f_1(z_2|z_1)&=\int_0^{\infty}x^{\gamma-1}[1+z_1^{\delta}(\alpha-1)
x^{\delta}]^{-\frac{1}{\alpha-1}}
{\rm e}^{-z_2^{\rho}x^{-\rho}}{\rm d}x\\
&\alpha>1,z_1>0,z_2>0,\delta>0,\rho>0. \mbox{  Put
$y=\frac{1}{x}$}\nonumber\\
f_1(z_1|z_2)&=\int_0^{\infty}y^{-\gamma-1}[1+z_1^{\delta}(\alpha-1)y^{-\delta}
]^{-\frac{1}{\alpha-1}}{\rm
e}^{-z_2^{\rho}y^{\rho}}{\rm d}y\\
&\alpha>1,z_1>0,z_2>0,\delta>0,\rho>0. \mbox{ Let $\alpha\rightarrow
1$ in (1)}\nonumber\\
f_2(z_2|z_1)&=\int_0^{\infty}x^{\gamma-1}{\rm
e}^{-z_1^{\delta}x^{\delta}}[1+z_2^{\rho}(\beta-1)x^{-\rho}]^{-\frac{1}{\beta-1}}{\rm
d}x\\
&\beta>1,z_1>0,z_2>0, \delta>0,\rho>0.\nonumber\\
f_2(z_1|z_2)&=\int_0^{\infty}x^{-\gamma-1}{\rm
e}^{-z_1^{\delta}x^{-\delta}}[1+z_2^{\rho}(\beta-1)x^{\rho}]^{-\frac{1}{\beta-1}}{\rm
d}x\\
&\beta>1,z_1>0,z_2>0,\delta>0,\rho>0. \mbox{ Take $\alpha\rightarrow
1, \beta\rightarrow 1$ in (1)}\nonumber\\
f_3(z_2|z_1)&=\int_0^{\infty}x^{\gamma-1}{\rm
e}^{-z_1^{\delta}x^{\delta}-z_2^{\rho}x^{-\rho}}{\rm d}x\\
&z_1>0, z_2>0, \delta>0, \rho>0.\nonumber\\
f_3(z_1|z_2)&=\int_0^{\infty}x^{-\gamma-1}{\rm
e}^{-z_1^{\delta}x^{-\delta}-z_2^{\rho}x^{\rho}}{\rm d}x\\
&z_1>0, z_2>0, \delta>0, \rho>0.\nonumber
\end{align}
\vskip.2cm In all the integrals considered so far, we had
one pathway factor containing $x^{\delta}$ and another
pathway factor containing $x^{-\rho}$, where both the
parameters $\delta>0$ and $\rho>0$, in the integrand. Also
the integrand consisted of non-negative integrable
functions and hence one could make statistical densities
out of them. In statistical terms, all the integrals
discussed so far will correspond to the density of
$u=x_1x_2$ where $x_1$ and $x_2$ are real scalar random
variables, which are statistically independently
distributed. \vskip.2cm Now we will consider a class of
integrals where the integrand consists of two pathway
factors where both contain powers of $x$ of the form
$x^{\delta}$ and $x^{\rho}$ with both $\delta$ and $\rho$
positive. Such integrals will lead to integrals of the
following forms in the limits when the pathway parameters
$\alpha$ and $\beta$ go to $1$:
\begin{equation*}
\int_0^{\infty}x^{\gamma}{\rm
e}^{-ax^{\delta}-bx^{\rho}},
\end{equation*}
$a>0,b>0,\delta>0,\rho>0$. Observe that the evaluation of
such an integral provides a method of evaluating Laplace
transform of generalized gamma densities by taking one of
the exponents $\delta$ or $\rho$ as unity. Consider the
integral
\begin{equation}
I_4=\int_0^{\infty}x^{\gamma}[1+z_1^{\delta}(\alpha
-1)x^{\delta}]^{-{{1}\over{\alpha-1}}}[1+z_2^{\rho}(\beta-1)x^{\rho}]^{-{{1}\over{\beta-1}}}{\rm
d}x,
\end{equation}
$\alpha>1,\beta>1,z_1>0,z_2>0,\delta>0,\rho>0$. Since the
integrand consists of positive integrable functions, from a
statistical point of view, the integral $I_4$ can be looked
upon as the density of $u=\frac{x_1}{x_2}$, where $x_1$ and
$x_2$ are real scalar random variables which are
independently distributed or it can be looked upon as a
convolution integral of the type
\begin{equation}
\int_0^{\infty}vf_1(uv)f_2(v){\rm d}v
\end{equation}
Let us take
\begin{align*}
f_1(x_1)&=c_1[1+(\alpha-1)x_1^{\delta}]^{-{{1}\over{\alpha-1}}},
u=z_1\\
f_2(x_2)&=c_2x^{\gamma-1}[1+z_2^{\rho}(\beta-1)x_2^{\rho}]^{-{{1}\over{\beta-1}}}\\
\intertext{Taking the Mellin transforms and writing as
expected values}\\
E(x_1)^{s-1}&=c_1\int_0^{\infty}x_1^{s-1}[1+(\alpha-1)x_1^{\delta}]^{-{{1}\over{\alpha-1}}}{\rm
d}x_1\\
&=\frac{c_1}{\delta(\alpha-1)^{\frac{s}{\delta}}}\frac{\Gamma\left(\frac{s}{\delta}\right)\Gamma\left(\frac{1}{\alpha-1}-\frac{s}{\delta}\right)}{\Gamma\left(\frac{1}{\alpha-1}\right)},\Re(s)>0,\Re(\frac{1}{\alpha-1}-\frac{s}{\delta})>0\\
E(x_2^{1-s})&=c_2\int_0^{\infty}x_2^{\gamma-s}[1+z_2^{\rho}(\beta-1)x_2^{\rho}]^{-\frac{1}{\beta-1}}\\
&=\frac{c_2}{\rho[z_2^{\rho}(\beta-1)]^{\frac{\gamma-s+1}{\rho}}}\frac{\left(\frac{\gamma-s+1}{\rho}\right)\Gamma\left(\frac{1}{\beta-1}-\frac{\gamma-s+1}{\rho}\right)}{\Gamma\left(\frac{1}{\beta-1}\right)}\\
&\Re(\gamma-s+1)>0,
\Re(\frac{1}{\beta-1}-\frac{\gamma-s+1}{\rho})>0.\\
\intertext{Therefore the density of $u=\frac{x_1}{x_2}$ is
given by}\\
g(u)&=\frac{c_1c_2}{\delta\rho[z_2^{\rho}(\beta-1)]^{\frac{\gamma+1}{\rho}}}\frac{1}{2\pi
i}\int_L\frac{\Gamma\left(\frac{\gamma+1}{\rho}-\frac{s}{\rho}\right)\Gamma\left(\frac{1}{\beta-1}-\frac{\gamma+1}{\rho}+\frac{s}{\rho}\right)}{\Gamma\left(\frac{1}{\alpha-1}\right)\Gamma\left(\frac{1}{\beta-1}\right)}\\
&\times
\Gamma\left(\frac{s}{\delta}\right)\Gamma\left(\frac{1}{\alpha-1}-\frac{s}{\delta}\right)\left[\frac{z_2(\alpha-1)^{\frac{1}{\delta}}}{z_1(\beta-1)^{\frac{1}{\rho}}}\right]^{-s}{\rm
d}s\\
&=\frac{c_1c_2}{\delta\rho[z_2^{\rho}(\beta-1)]^{\frac{\gamma+1}{\rho}}\Gamma\left(\frac{1}{\alpha-1}\right)\Gamma\left(\frac{1}{\beta-1}\right)}\\
&\times
H_{2,2}^{2,2}\left[\frac{z_2(\alpha-1)^{\frac{1}{\delta}}}{z_1(\beta-1)^{\frac{1}{\rho}}}\bigg\vert_{\left(0,\frac{1}{\delta}\right),\left(\frac{1}{\beta-1}-\frac{\gamma+1}{\rho},\frac{1}{\rho}\right)}^{\left(1-\frac{\gamma+1}{\rho},\frac{1}{\rho}\right),\left(1-\frac{1}{\alpha-1},\frac{1}{\delta}\right)}\right].\\
\intertext{Therefore}\\
I_4&=\int_0^{\infty}x^{\gamma}[1+z_1^{\delta}(\alpha-1)x^{\delta}]^{-\frac{1}{\alpha-1}}[1+z_2^{\rho}(\beta-1)x^{\rho}]^{-\frac{1}{\beta-1}}{\rm
d}x,\\
&~~~~~~\alpha>1,\beta>1,\delta>0,\rho>0\\
&=\frac{1}{\delta\rho[z_2^{\rho}(\beta-1)]^{\frac{\gamma+1}{\rho}}\Gamma\left(\frac{1}{\alpha-1}\right)\Gamma\left(\frac{1}{\beta-1}\right)}\\
&\times
H_{2,2}^{2,2}\left[\frac{z_2(\alpha-1)^{\frac{1}{\delta}}}{z_1(\beta-1)^{\frac{1}{\rho}}}\bigg\vert_{\left(0,\frac{1}{\delta}\right),\left(\frac{1}{\beta-1}-\frac{\gamma+1}{\rho},\frac{1}{\rho}\right)}^{\left(1-\frac{\gamma+1}{\rho},\frac{1}{\rho}\right),\left(1-\frac{1}{\alpha-1},\frac{1}{\delta}\right)}\right]
\intertext{Now by putting $y=\frac{1}{x}$ we can get an
associated integral}\\
I_4&=\int_0^{\infty}y^{-\gamma-2}[1+z_1^{\delta}(\alpha-1)y^{-\delta}]^{-\frac{1}{\alpha-1}}
[1+z_2^{\rho}(\beta-1)y^{-\rho}]^{-\frac{1}{\beta-1}}{\rm
d}y.
\end{align*}
Now, we can look at various special cases of
$\lim_{\alpha\rightarrow 1}$ or $\lim_{\beta\rightarrow 1}$
or $\lim_{\alpha\rightarrow 1,\beta\rightarrow 1}$. These
lead to some interesting special cases.
\begin{align*}
I_{4.1}&=\lim_{\alpha\rightarrow 1_{+}}I_4\\
&=\int_0^{\infty}x^{\gamma}{\rm
e}^{-z_1^{\delta}x^{\delta}}[1+z_2^{\rho}(\beta-1)x^{\rho}]^{-\frac{1}{\beta-1}}{\rm
d}x\\
&=\frac{1}{\rho\delta[z_2^{\rho}(\beta-1)]^{\frac{\gamma+1}{\rho}}
\Gamma\left(\frac{1}{\beta-1}\right)}H_{1,2}^{2,1}\left[\frac{z_2}{z_1(\beta-1)^{\frac{1}{\rho}}}
\bigg\vert_{\left(0,\frac{1}{\delta}\right),\left(\frac{1}{\beta-1}-\frac{\gamma+1}{\rho},\frac{1}{\rho}\right)}^{\left(1-\frac{\gamma+1}{\rho},\frac{1}{\rho}\right)}\right].\\
I_{4.2}&=\lim_{\beta\rightarrow 1_{+}}I_4\cr
&=\int_0^{\infty}x^{\gamma}[1+z_1^{\delta}(\alpha-1)x^{\delta}]^{-\frac{1}{\alpha-1}}{\rm
e}^{-z_2^{\rho}x^{\rho}}{\rm d}x\\
&=\frac{1}{\rho\delta
z_2^{\gamma+1}\Gamma\left(\frac{1}{\alpha-1}\right)}
H_{2,1}^{1,2}\left[\frac{z_2(\alpha-1)^{\frac{1}{\delta}}}{z_1}
\bigg\vert_{\left(0,\frac{1}{\delta}\right)}^{\left(1-\frac{\gamma+1}{\rho},
\frac{1}{\rho}\right),\left(1-\frac{1}{\alpha-1},\frac{1}{\delta}\right)}\right]\\
I_{4.3}&=\lim_{\alpha\rightarrow 1,\beta\rightarrow 1}I_4\\
&=\int_0^{\infty}x^{\gamma}{\rm
e}^{-(z_1x)^{\delta}-(z_2x)^{\rho}}{\rm d}x\\
&=\frac{1}{\rho\delta z_2^{\frac{\gamma+1}{\rho}}}\\
&\times H_{1,1}^{1,1}\left[\frac{z_2}{z_1}
\bigg\vert_{\left(0,\frac{1}{\delta}\right)}^{\left(1-\frac{\gamma+1}{\rho},
\frac{1}{\rho}\right)}\right]\\
&=\int_0^{\infty}x^{-\gamma-2}{\rm
e}^{-z_1^{\delta}x^{-\delta}-z_2^{\rho}x^{-\rho}}{\rm d}x.
\end{align*}
When $\alpha<1$ and $\beta<1$ also we can obtain
corresponding integrals, which are finite range integrals,
by going through parallel procedure. In this case the limit
of integration will be $0<x<\max\{\epsilon_1,\epsilon_2\}$
where
$\epsilon_1=[z_1^{\delta}(\alpha-1)]^{-\frac{1}{\delta}}$
and $\epsilon_2=[z_2^{\rho}(\beta-1)]^{-\frac{1}{\rho}}$.
The details of the integrals will not be listed here in
order to save space.
 \vskip.3cm \noindent
 {\bf 1.2.\hskip.3cm Case of $\alpha<1$, or
$\beta <1$.} \vskip.3cm When $\alpha<1$, writing
$\alpha-1=-(1-\alpha)$ we can define the function
\begin{equation}
g_1(x)=x^{\gamma}[1+z_1^{\delta}(\alpha-1)x^{\delta}]^{-\frac{1}{\alpha-1}}
=x^{\gamma}[1-z_1^{\delta}(1-\alpha)x^{\delta}]^{\frac{1}{1-\alpha}},\alpha<1
\end{equation}
for $[1-z_1^{\delta}(1-\alpha)x^{\delta}]>0, \alpha<1\Rightarrow
x<\frac{1}{z_1(1-\alpha)^{\frac{1}{\delta}}}$ and $g_1(x)=0$
elsewhere. In this case the Mellin transform of $g_1(x)$ is the
following:
\begin{align}
h_1(s)&=\int_0^{\infty} x^{s-1}g_1(x){\rm
d}x=\int_0^{\frac{1}{z_1(1-\alpha)^{\frac{1}{\delta}}}}x^{\gamma+s-1}
[1-z_1^{\delta}(1-\alpha)x^{\delta}]^{\frac{1}{1-\alpha}}{\rm
d}x\\
&=\frac{1}{\delta
[z_1(1-\alpha)^{\frac{1}{\delta}}]^{\gamma+s}}\frac{\Gamma(\frac{\gamma+s}{\delta})
\Gamma(\frac{1}{1-\alpha}+1)}{\Gamma(\frac{1}{1-\alpha}+1+\frac{\gamma+s}{\delta})},
\Re(\gamma+s)>0,\alpha<1,\delta>0.
\end{align}
Then the Mellin transform of $f(z_2|z_1)$ for $\alpha<1,\beta>1$ is
given by
\begin{align}
M_{z_2|z_1}(s)&=\frac{\Gamma(\frac{1}{1-\alpha}+1)}{\delta\rho
z_2^sz_1^{\gamma+s}(\beta-1)^{\frac{s}{\rho}}(1-\alpha)^{\frac{\gamma+s}{\delta}}}
\frac{\Gamma(\frac{\gamma+s}{\delta})}{\Gamma(\frac{\gamma+s}{\delta}
+\frac{1}{1-\alpha}+1)}\frac{\Gamma(\frac{s}{\rho})\Gamma(\frac{1}{\beta-1}
-\frac{s}{\rho})}{\Gamma(\frac{1}{\beta-1)}},\\
&\Re(\gamma+s)>0, \Re(s)>0,
\Re\left(\frac{1}{\beta-1}-\frac{s}{\rho}\right)>0.\notag
\end{align}
Hence the inverse Mellin transform for $\alpha<1, \beta>1$ is
\begin{align}
f(z_2|z_1)&=\frac{\Gamma(\frac{1}{1-\alpha}+1)}{\delta\rho
z_1^{\gamma}(1-\alpha)^{\frac{\gamma}{\delta}}\Gamma(\frac{1}{\beta-1})}\nonumber\\
&\times
H_{2,2}^{2,1}\left[z_1z_2(1-\alpha)^{\frac{1}{\delta}}(\beta-1)^{\frac{1}{\rho}}
\big\vert_{(0,\frac{1}{\rho}),
(\frac{\gamma}{\delta},\frac{1}{\delta})}^{(1-\frac{1}{\beta-1},\frac{1}{\rho}),
(1+\frac{1}{1-\alpha}+\frac{\gamma}{\delta},\frac{1}{\delta})}\right]\\
\lim_{\beta\rightarrow
1}f(z_2|z_1)&=\frac{\Gamma(\frac{1}{1-\alpha}+1)}{\rho\delta
z_1^{\gamma}(1-\alpha)^{\frac{\gamma}{\delta}}}
H_{1,2}^{2,0}\left[z_1z_2(1-\alpha)^{\frac{1}{\delta}}
\big\vert_{(0,\frac{1}{\delta}), (\frac{\gamma}{\delta},
\frac{1}{\delta})}^{(1+\frac{1}{1-\alpha}+\frac{\gamma}{\delta},\frac{1}{\delta})}\right]\\
\lim_{\alpha\rightarrow
1}f(z_2|z_1)&=\frac{1}{\rho\delta\Gamma(\frac{1}{\beta-1})z_1^{\gamma}}
H_{1,2}^{2,1}\left[z_1z_2(\beta-1)^{\frac{1}{\rho}}\big\vert_{(0,\frac{1}{\rho}),
(\frac{\gamma}{\delta},
\frac{1}{\delta})}^{(1-\frac{1}{\beta-1},\frac{1}{\rho})}\right]\\
\lim_{\alpha\rightarrow 1,\beta\rightarrow
1}f(z_2|z_1)&=\frac{1}{\rho\delta z_1^{\gamma}}
H_{0,2}^{2,0}\left[z_1z_2\big\vert_{(0,\frac{1}{\rho}),(\frac{\gamma}{\delta},
\frac{1}{\delta})}\right].
\end{align}
\vskip.2cm In $f(z_2|z_1)$ if $\beta<1$ we may write
$\beta-1=-(1-\beta)$, and if we assume
$[1-z_2^{\rho}(1-\beta)x^{-\rho}]^{\frac{1}{1-\beta}}>0\Rightarrow
x>z_2(1-\beta)^{\frac{1}{\rho}}$ then also the corresponding
integrals can be evaluated as H-functions. But if $\alpha<1$ and
$\beta<1$ then from the conditions
\begin{equation*}
1-z_1^{\delta}(1-\alpha)x^{\delta}>0\Rightarrow
x<\frac{1}{z_1(1-\alpha)^{\frac{1}{\delta}}}\mbox{  and  }
1-z_2^{\rho}(1-\beta)x^{-\rho}>0\Rightarrow
x>z_2(1-\beta)^{\frac{1}{\rho}}
\end{equation*}
and the resulting integral may be zero. Hence, except this case of
$\alpha<1$ and $\beta<1$ all other cases: $\alpha>1,\beta>1;
\alpha<1,\beta>1;\alpha>1,\beta<1$ can be given meaningful
interpretations as H-functions. Further, all these situations can be
connected to practical problems. A few such practical situations
will be considered next. \vskip.3cm \noindent {\bf 2.\hskip.3cm Specific Applications} 
\vskip.3cm \noindent {\bf 2.1.\hskip.3cm
Kr\"atzel Integral} \vskip.3cm For $\delta =1, z_2^{\rho}=z, z_1=1$
in $f_3(z_2|z_1)$ gives the Kr\"atzel integral
\begin{equation}
f_3(z_2|z_1)=\int_0^{\infty}x^{\gamma-1}{\rm e}^{-x-z x^{-\rho}}{\rm
d}x
\end{equation}
which was studied in detail by Kr\"atzel (1979). Hence $f_3$ can be
considered as generalization of Kr\"atzel integral. An additional
property that can be seen from Kr\"atzel integral as $f_3$ is that
it can be written as a H-function of the  type
$H_{0,2}^{2,0}(\cdot)$. Hence all the properties of H-function can
now be made use of to study this integral further. \vskip.3cm
\noindent
{\bf 2.2.\hskip.3cm Inverse Gaussian Density in
Statistics} \vskip.3cm Inverse Gaussian density is a popular
density, which is used in many disciplines including stochastic
processes. One form of the density is the following (Mathai, 1993,
page 33):
\begin{equation}
f(x)=c~x^{-\frac{3}{2}}{\rm
e}^{-\frac{\lambda}{2}(\frac{x}{\mu^2}+\frac{1}{x})}, \mu\ne 0, x>0,
\lambda>0
\end{equation}
where $c=\pi^{-\frac{1}{2}}{\rm e}^{\frac{\lambda}{|\mu|}}$.
Comparing this with our case $f_3(z_1|z_2)$ we see that the inverse
Gaussian density is the integrand in $f_3(z_1|z_2)$ for
$\gamma=\frac{1}{2}, \rho =1,
z_2=\frac{\lambda}{2}(\frac{1}{\mu^2}),
\delta=1,z_1=\frac{\lambda}{2}$. Hence $f_3$ can be used directly to
evaluate the moments or Mellin transform in inverse Gaussian
density.
\vskip.3cm \noindent
{\bf 2.3.\hskip.3cm Reaction Rate
Probability Integral in Astrophysics} \vskip.3cm In a series of
papers Haubold and Mathai studied modifications to Maxwell-Boltzmann
theory of reaction rates, a summary is given in Mathai and Haubold
(1988) and Mathai and Haubold (2008). The basic reaction rate probability integral that appears
there is the following:
\begin{equation}
I_1=\int_0^{\infty}x^{\gamma-1}{\rm e}^{-ax-zx^{-\frac{1}{2}}}{\rm
d}x.
\end{equation}
This is the case in the non-resonant case of nuclear reactions.
Compare integral $I_1$ with $f_3(z_2|z_1)$. The reaction rate
probability integral $I_1$ is $f_3(z_2|z_1)$ for $\delta=1,
\rho=\frac{1}{2},z_2^{\frac{1}{2}}=z$. The basic integral $I_1$ is
generalized in many different forms for various situations of
resonant and non-resonant cases of reactions, depletion of high
energy tail, cut off of the high energy tail and so on. Dozens of
published papers are there in this area. \vskip.3cm \noindent {\bf
2.4.\hskip.3cm Tsallis Statistics and Superstatistics} \vskip.3cm 
Tsallis statistics is of the following form:
\begin{equation}
f_x(x)=c_1 [1+(\alpha-1)x]^{-\frac{1}{\alpha -1}}.
\end{equation}
Compare $f_x(x)$ with the integrand in (1). For $z_2=0, \delta=1,
\gamma=1$ the integrand in (1) agrees with Tsallis statistics
$f_x(x)$ given above. The three different forms of Tsallis
statistics are available from $f_x(x)$ for
$\alpha>1,\alpha<1,\alpha\rightarrow 1$. The starting paper in
non-extensive statistical mechanics may be seen from Tsallis (2009).
But the integrand in (1) with $z_2=0, z_1=1, \alpha>1$ is the
superstatistics of Beck and Cohen, see for example Beck and Cohen
(2003), Beck (2006). In statistical language, this superstatistics
is the posterior density in a generalized gamma case when the scale
parameter has a prior density belonging to the same class of
generalized gamma density.
\vskip.3cm \noindent {\bf 2.5.\hskip.3cm
Pathway Model} \vskip.3cm Mathai (2005) considered a rectangular
matrix-variate function in the real case from where one can obtain
almost all matrix-variate densities in current use in statistical
and other disciplines. The corresponding version when the elements
are in the complex domain is given in Mathai and Provost (2006). For
the real scalar case the function is of the following form:
\begin{equation}
f(x)=c^{*}|x|^{\gamma}[1-a(1-\alpha)|x|^{\delta}]^{\frac{\eta}{1-\alpha}}
\end{equation}
for $-\infty<x<\infty, a>0, \eta>0,\delta>0$ and $c^{*}$ is the
normalizing constant. Here $f(x)$ for $\alpha<1$ stays in the
generalized type-1 beta family when
$[1-a(1-\alpha)|x|^{\delta}]^{\frac{\eta}{1-\alpha}}>0$. When
$\alpha>1$ the function switches into a generalized type-2 beta
family and when $\alpha\rightarrow 1$ it goes into a generalized
gamma family of functions. Here $\alpha$ behaves as a pathway
parameter and hence the model is called a pathway model. Observe
that the integrand in (1) is a product of two such pathway functions
so that the corresponding integral is more versatile than a pathway
model. Thus for $z_2=0$ in (1) the integrand produces the pathway
model of Mathai (2005).\vskip.3cm \noindent
\begin{center} {\bf
Acknowledgement}
\end{center}
\vskip.3cm The author would like to acknowledge with thanks the
financial assistance from the Department of Science and Technology,
Government of India, New Delhi, under Project No. SR/S4/MS:287/05
\vskip.3cm \noindent
\begin{center}
{\bf
References}
\end{center}
Beck,C. (2006): Stretched exponentials from sueprstatistics, {\it Physica A} {\bf 365}, 96-101.\\
\vskip.2cm \noindent Beck, C. and Cohen, E.G.D. (2003): Superstatistics, {\it Physica A} {\bf 322}, 267-275.\\
\vskip.2cm \noindent Kr\"atzel, E. (1979): Integral transformations of Bessel type. In {\it Generalized Functions and Operational Calculus}, Proc. Conf. Varna, 1975, Bulg. Acad. Sci. Sofia, 148-165.\\
 \vskip.2cm \noindent Mathai, A.M. (1993): {\it A Handbook of Generalized Special Functions for Statistical and Physical Sciences}, Oxford University Press, Oxford.\\
\vskip.2cm \noindent Mathai, A.M. (2005): A pathway to matrix-variate gamma and normal densities. {\it Linear Algebra and Its Applications}, {\bf 396}, 317-328.\\
\vskip.2cm \noindent Mathai, A.M. and Haubold, H.J. (1988): {\it Modern Problems in Nuclear and Neutrino Astrophysics}, Akademie-Verlag, Berlin\\
\vskip.2cm \noindent Mathai, A.M. and Haubold, H.J. (2008): {\it Special Functions for Applied Scientists}, Springer, New Yoek\\ 
\vskip.2cm \noindent Mathai, A.M.  and Provost, S.B. (2006): Some complex matrix-variate statistical distributions on rectangular matrices. {\it Linear Algebra and Its Applications}, {\bf 410}, 198-216.\\
\vskip.2cm \noindent Mathai, A.M. and Saxena, R.K. (1978): {\it The H-function with Applications in Statistics and Other Disciplines}, Wiley Halsted, New York and Wiley Eastern, New Delhi.\\
\vskip.2cm \noindent Mathai, A.M., Saxena, R.K., and Haubold, H.J. (2010): {\it The H-Function: Theory and Applications}, Springer, New York.\\
 \vskip.2cm \noindent Tsallis, C. (2009): {\it Introduction to Nonextensive Statistical Mechanics: Approaching a Complex World}, Springer, New York.\\
\end{document}